\documentclass[12pt]{article}
\usepackage{graphics,color}
\begin{document}
\thispagestyle{empty}
\noindent\
\begin{center}
\large \bf Flavor Mixing of Quarks and a NewTexture
\end{center}
\hfill
 \vspace*{1cm}
\noindent
\begin{center}
{\bf Harald Fritzsch}\\
Department f\"ur Physik\\ 
Universit\"at M\"unchen\\
Theresienstra{\ss}e 37\\
D-80333 M\"unchen, 
Germany
\vspace*{0.5cm}
\end{center}

\begin{abstract}

We discuss a new  mass matrix with specific texture zeros for the quarks. The three flavor mixing angles for the quarks are functions of the quark masses and can be calculated.
The following ratios among  CKM matrix elements are given by ratios of quark masses: 

 $|V^{}_{td}/V^{}_{ts}| \simeq \sqrt{m^{}_d/m^{}_s}$
and $|V^{}_{ub}/V^{}_{cb}| \simeq \sqrt{m^{}_u/m^{}_c}$.

Also we can calculate two CKM matrix elements:

$|V^{}_{cb}| \simeq |V^{}_{ts}| \simeq 2\left(m^{}_s/m^{}_b\right)$. 

This relation as well as the relation $|V^{}_{td}/V^{}_{ts}| \simeq \sqrt{m^{}_d/m^{}_s}$ are in good agreement with the
experimental data. There is a problem with the relation $|V^{}_{ub}/V^{}_{cb}| \simeq \sqrt{m^{}_u/m^{}_c}$, probably due to wrong estimates of the quark masses  $m^{}_u$ and  $m^{}_c$.

\end{abstract}

\newpage

The flavor mixing of the quarks and leptons is still a mystery in physics, and thus far it is not possible to calculate the flavor mixing angles. 
These angles have been measured in many experiments. For the six quarks the flavor mixing is described by the CKM matrix:

\begin{eqnarray}
V^{}_{CKM}= \left( \matrix{ V^{}_{ud}
& V^{}_{us} & V^{}_{ub} \cr
V^{}_{cd} &
V^{}_{cs} &
V^{}_{cb} \cr V^{}_{td} & V^{}_{ts} & V^{}_{tb} \cr} \right). 
\end{eqnarray} 

The absolute values of the nine matrix elements have been measured to a high
degree of accuracy, as given by the Particle Data group \cite{1}:\\

\begin{eqnarray}
\left|V^{}_{\rm CKM}\right| =
\left( \matrix{ 0.97401 \pm 0.00011 & 0.22650 \pm 0.00048
& 0.00361^{+0.00011}_{-0.00009} \cr 0.22636 \pm 0.00048 & 0.97320 \pm 0.00011
& 0.04053^{+0.00083}_{-0.00061} \cr
0.00854^{+0.00023}_{-0.00016} & 0.03978^{+0.00082}_{-0.00060}
& 0.999172^{+0.000024}_{-0.000035} \cr} \right) \;.
\end{eqnarray}\\

The matrix elements $|V^{}_{us}|$ and $|V^{}_{cd}|$ are approximately equal, also the matrix elements 
$|V^{}_{cb}|$ and $|V^{}_{ts}|$. The following two ratios of the CKM matrix elements are particularly
interesting and will be discussed later in more detail:

\begin{eqnarray}
\left|\frac{V^{}_{ub}}{V^{}_{cb}}\right| = 0.0891^{+0.0041}_{-0.0040} \; , \quad
\left|\frac{V^{}_{td}}{V^{}_{ts}}\right| = 0.2147^{+0.0092}_{-0.0083} \; .
\end{eqnarray}
\\
There are several ways to describe the CKM-matrix in terms of three angles and one phase parameter.
 I prefer the parametrization, which Z. Xing and I introduced years ago \cite{2}, given by the angles
 $\theta_u$, $\theta_d$, $\theta$ and a phase parameter $\phi$, which describes CP violation:

\begin{eqnarray}
V^{}_{\rm CKM} \hspace{-0.2cm} & = & \hspace{-0.2cm} \left( \matrix{ c^{}_{\rm u}
& s^{}_{\rm u} & 0 \cr
-s^{}_{\rm u} &
c^{}_{\rm u} &
0 \cr 0 & 0 & 1 \cr} \right)\times \left( \matrix{ e^{-{\rm i}\phi}
& 0 & 0 \cr
0 &
c &
s \cr 0 & -s & c \cr} \right) \times \left( \matrix{ c^{}_{\rm }
& - s^{}_{\rm d} & 0 \cr
s^{}_{\rm d} &
c^{}_{\rm d} &
0 \cr 0 & 0 & 1 \cr} \right)
\nonumber \\
\hspace{-0.2cm} & = & \hspace{-0.2cm}
\left(\matrix{ s^{}_{\rm u} s^{}_{\rm d} c + c^{}_{\rm u} c^{}_{\rm d}
e^{-{\rm i} \phi} & s^{}_{\rm u} c^{}_{\rm d} c - c^{}_{\rm u} s^{}_{\rm d}
e^{-{\rm i} \phi} & s^{}_{\rm u} s \cr
c^{}_{\rm u} s^{}_{\rm d} c -
s^{}_{\rm u} c^{}_{\rm d} e^{-{\rm i}\phi} & c^{}_{\rm u} c^{}_{\rm d} c
+ s^{}_{\rm u} s^{}_{\rm d} e^{-{\rm i}\phi} & c^{}_{\rm u} s
\cr - s^{}_{\rm d} s   & - c^{}_{\rm d} s   & c \cr}\right) \; .
\end{eqnarray}

Here we used the short notation: $c^{}_{u,d} \sim \cos\theta^{}_{u,d}$, $s^{~}_{u,d}
\sim \sin\theta^{}_{u,d}$, $c \sim \cos\theta$ and $s \sim
\sin\theta$.\\

These four parameters, three angles and one phase parameter, can be determined from the experimental data, given in eqs. (2) and (3). Their central values are:

\begin{eqnarray}
\theta_u \simeq 5.0^\circ \;,  \quad \theta_d \simeq 12.3^\circ \; ,
\quad \theta \simeq 2.4^\circ \; , \quad \phi \simeq 90^\circ \; .
\end{eqnarray}

Relations between the quark masses and the mixing angles can be derived, if the quark mass matrices have "texture zeros", 
as shown by S. Weinberg and me in 1977 \cite{3}.\\

 In 1978 I studied the flavor mixing of six quarks. Here I used a mass matrix 
with four "texture zeros" \cite{4}: 

\begin{eqnarray}
M_{\rm q} = \left( \matrix{ 0 & A_{\rm q} & 0 \cr A^*_{\rm q} & 0 & B_{\rm q} \cr
0 & B^*_{\rm q} & C_{\rm q} \cr} \right) \;.
\end{eqnarray}
Here q stands for "u" (the "upsector") or "d" ( the "down sector"). This specific mass matrix has been studied later by many physicists.\\

Now the two mass matrices are diagonalized by the unitary transformations 
$V^\dagger_{\rm u}
M^{}_{\rm u} V^{}_{\rm u} = {\rm diag}\{m^{}_u, - m^{}_c,
m^{}_t\}$ and  $V^\dagger_{\rm d}
M^{}_{\rm d} V^{}_{\rm d} = {\rm diag}\{m^{}_d, - m^{}_s,
m^{}_b\}$.
\\

The CKM matrix is given by the product of the two unitary matrices:  $V \equiv V^\dagger_{\rm u}
V^{}_{\rm d}$. Thus one can calculate 
the CKM mixing angles $\theta_u$, $\theta_d$ and $\theta$ as functions of the quark masses. In a good approximation we find: 

\begin{equation}
\tan\theta^{}_d \simeq
|V^{}_{td}/V^{}_{ts}| \simeq \sqrt{m^{}_d/m^{}_s},\hspace*{1cm}
\tan\theta^{}_u \simeq|V^{}_{ub}/V^{}_{cb}|\simeq
\sqrt{m^{}_u/m^{}_c}.
\end{equation}

\begin{equation}
\sin\theta \simeq
|V^{}_{cb}| \simeq |\sqrt{m^{}_s/m^{}_b} -e^{{\rm i} \varphi^{}}\sqrt{m^{}_c/m^{}_t}|.
\end{equation}\\  

In the last relation a phase parameter appears - the phase difference between the off-diagonal matrix elments B and B*.\\

The values of the six quark masses have been measured in many experiments. These values depend on the energy scale. They have recently been
calculated at the energy scale, given by the mass
 of the Z-boson: $M^{}_Z \simeq 91.2 ~{\rm GeV} $ ( \cite{5}, \cite{6} ):\\

$m^{}_u = 1.24 \pm 0.22 ~{\rm MeV} ,
m^{}_c = 0.62 \pm 0.02 ~{\rm GeV} ,
m^{}_t = 168.26 \pm 0.75 ~{\rm GeV}.$\\

$m^{}_d = 2.69 \pm 0.19 ~{\rm MeV}, 
m^{}_s = 53.5 \pm 4.6 ~{\rm MeV}, 
m^{}_b = 2.86 \pm 0.03 ~{\rm GeV}.$\\

Thus the relative uncertainty of a quark mass is larger for the small quark masses - 
the mass of the t-quark has the smallest relative error - only about $0.5\%$. For the u-quark 
the relative error is about 17 percent. \\
\\
Using these quark masses and eq.(8),  one finds, if the phase parameter is zero: 

\begin{equation}
|V^{}_{cb}| \simeq  0.076.
\end{equation}
But the eperiments give (see eq. (2)): 

\begin{equation}
|V^{}_{cb}| \simeq  0.041.
\end{equation}
If the phase parameter is not zero, $|V^{}_{cb}|$ is even larger. Thus the mass matrix  with four texture zeros, given in eq. (6), is not correct. \\

I was searching for a mass matrix with a specific texture, which allows to calculate $V^{}_{cb}$ and $V^{}_{ts}$ correctly. 
Now I found the following mass matrix with three texture zeros: 

\begin{eqnarray}
M_{\rm q}= \left( \matrix{ 0 & A & 0 \cr A^* & |B/2| & B \cr
0 & B^* & C \cr} \right) \;.
\end{eqnarray}\\

Although a nonzero (2,2) matrix element has been introduced  to modify my original texture (eq. (6)) ( see e.g.\cite{7,8}), it is the first time that the specific texture in eq. (11) is considered.\\

This matrix can be diagonalized by an orthogonal transformation, which is similar to the transformation, used for the mass matrix, given in eq. (6). Using the parameters of this transformation, one can calculate the mixing angles as 
functions of the quark masses: 

\begin{equation}
\tan\theta^{}_d \simeq
|V^{}_{td}/V^{}_{ts}| \simeq \sqrt{m^{}_d/m^{}_s},\hspace*{1cm}\\
\tan\theta^{}_u \simeq|V^{}_{ub}/V^{}_{cb}|\simeq
\sqrt{m^{}_u/m^{}_c}.
\end{equation}

These two relations did not change, if the mass matrix is changed from (6)  to (11). But the third relation changes: 

\begin{eqnarray}
\sin\theta \simeq|V_{cb}| \simeq |V_{ts}|  \simeq 2 \left|\frac{m^{}_s}{m^{}_b} -
e^{{\rm i} \varphi^{}} \frac{m^{}_c}{m^{}_t}\right| \simeq 2 \frac{m_s}{m_b} \;.
\end{eqnarray}

Here again a phase parameter $\varphi$ appears, since the mass matrices have complex off-diagonal matrixelements. However in a good approximation this phase can be neglected, since the mass ratio (${m_c}/{m_t})$
 is very small - it contributes only about  $0.4\%$. If the ratio (${m_c}/{m_t}$) is neglected, we obtain: 

\begin{eqnarray}
|V_{cb}|\simeq |V_{ts}|  \simeq 2 \frac{m_s}{m_b} \;.
\end{eqnarray}

Now we compare the experimental value of $|V^{}_{cb}|$ and the estimated value of ${m_s}$/${m_b}$: 
 \begin{eqnarray}
\left|V^{}_{cb}\right|: 0.0395 =>0.0415,
\end{eqnarray}
\begin{eqnarray}
2 \frac{m_s}{m_b}: 0.034 => 0.042.
\end{eqnarray}

In a good approximation both numbers are equal. Thus the third angle of the CKM matrix is also given by a ratio of quark masses: (${m_s} /{m_b}$). \\

Now we consider the ratio: 

\begin{eqnarray}
\left|\frac{V^{}_{td}}{V^{}_{ts}}\right| = 0.2147^{+0.0092}_{-0.0083} \; .
\end{eqnarray}
This ratio can be calculated - it is given by the ratio of $m^{}_d$/$m^{}_s$:

\begin{equation}
|V^{}_{td}/V^{}_{ts}| \simeq \sqrt{m^{}_d/m^{}_s} \simeq 0.224.
\end{equation}\\
Thus the experimental number and the calculated value agree very well. But there is a problem with the second relation: 

\begin{eqnarray}
&& \left|\frac{V^{}_{ub}}{V^{}_{cb}}\right| = 0.0891^{+0.0041}_{-0.0040} \; .
\end{eqnarray}
Again this ratio can be calculated. It is given by the ratio ($m^{}_u$/$m^{}_c$). If the quark masses, which 
are quoted above, are used, one finds: 

\begin{equation}
|V^{}_{ub}/V^{}_{cb}| \simeq \sqrt{m^{}_u/m^{}_c} \simeq 0.045.
\end{equation}
The experimental value is almost twice as large as the value, given by the ratio  ($m^{}_u$/$m^{}_c$), and the question arises: Is this mass ratio 
correct?\\ 

The ratio ($m^{}_d$/$m^{}_s$) can be calculated in chiral perturbation theory - thus it is better known 
than the quark masses $m^{}_d$ or $m^{}_s$.\\

However the ratio  ($m^{}_u$/$m^{}_c$) cannot be calculated in chiral perturbation theory. It can only be calculated by using the actual values of the 
quark masses. The errors of the quark masses, quoted above, are probably much larger than indicated. For example, the up quark mass could be 3.8 MeV and  the mass 
of the charm quark about 600 MeV. In this case we would obtain:

\begin{equation}
|V^{}_{ub}/V^{}_{cb}| \simeq \sqrt{m^{}_u/m^{}_c} \simeq 0.08.
\end{equation}

Thus there would be no problem with the experimental value.\\

It seems that the quark mass matrices with the texture, given in eq. (11), describe very well the flavor mixing of the quarks. The quark flavors mix, since the mass matrix of the quarks has this specific texture. 
This texture might be due to a new discrete symmetry, which is beyond the Standard Model of particle physics. A deeper understanding of this texture might open the road to the physics beyond the Standard Model. \\
\\

\

\end{document}